\documentclass[conference]{IEEEtran}

\usepackage{algorithm}
\usepackage{cite}
\usepackage{subfigure}
\usepackage{multirow}
\usepackage{url}
\usepackage{cite}
\usepackage{amsmath,amssymb,amsfonts}
\usepackage{graphicx}
\usepackage{textcomp}
\usepackage{algpseudocode} 
\usepackage[table]{xcolor}
\usepackage{makecell}
\usepackage{arydshln}
\usepackage[linkcolor=red,urlcolor=blue!70!black]{hyperref}

\begin{document}


\title{Graph Representation Learning with Massive Unlabeled Data for Rumor Detection}

\author{\textbf{Chaoqun Cui, Caiyan Jia$^{*}$} \\
\IEEEauthorblockA{\textit{School of Computer and Information Technology \& Beijing Key Lab of Traffic Data Analysis and Mining} \\
\textit{Beijing Jiaotong University}\\
Beijing 100044, China \\
\{21120341,cyjia\}@bjtu.edu.cn}\\
\small{$^{*}$Corresponding author}
}

\maketitle

\begin{abstract}

With the development of social media, rumors spread quickly, cause great harm to society and economy. Thereby, many effective rumor detection methods have been developed, among which the rumor propagation structure learning based methods are particularly effective compared to other methods. However, the existing methods still suffer from many issues including the difficulty to obtain large-scale labeled rumor datasets, which leads to the low generalization ability and the performance degeneration on new events since rumors are time-critical and usually appear with hot topics or newly emergent events. In order to solve the above problems, in this study, we used large-scale unlabeled topic datasets crawled from the social media platform Weibo and Twitter with claim propagation structure to improve the semantic learning ability of a graph reprentation learing model on various topics. We use three typical graph self-supervised methods, InfoGraph, JOAO and GraphMAE in two commonly used training strategies, to verify the performance of general graph semi-supervised methods in rumor detection tasks. In addition, for alleviating the time and topic difference between unlabeled topic data and rumor data, we also collected a rumor dataset covering a variety of topics over a decade (10-year ago from 2022) from the Weibo rumor-refuting platform. Our experiments show that these general graph self-supervised learning methods outperform previous methods specifically designed for rumor detection tasks and achieve good performance under few-shot conditions, demonstrating the better generalization ability with the help of our massive unlabeled topic dataset.

\end{abstract}


\section{INTRODUCTION}

\begin{figure}[!h]
  \centering
  \subfigure[Rumor]{\includegraphics[width=0.45\linewidth]{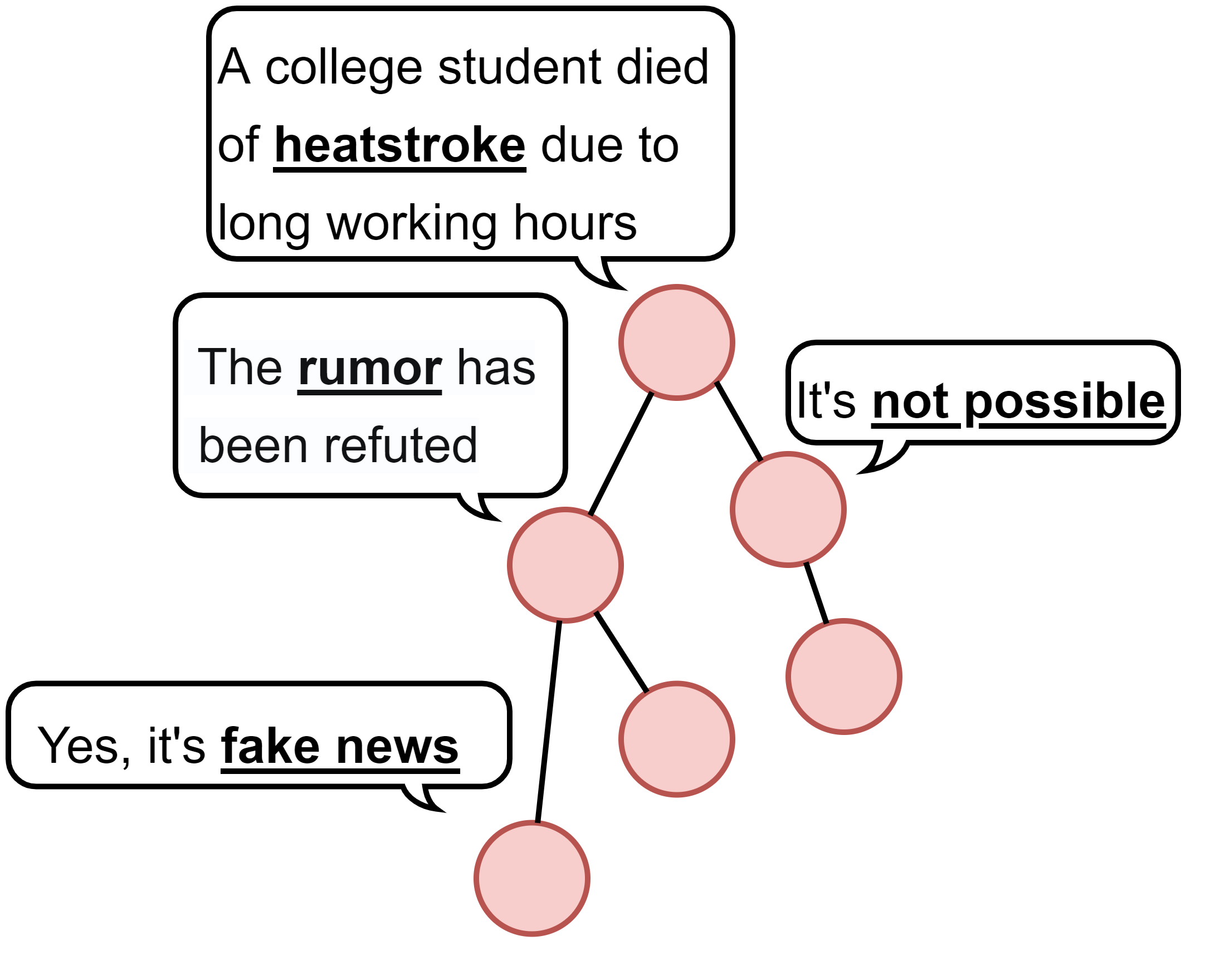}}
  \subfigure[Non-rumor]{\includegraphics[width=0.45\linewidth]{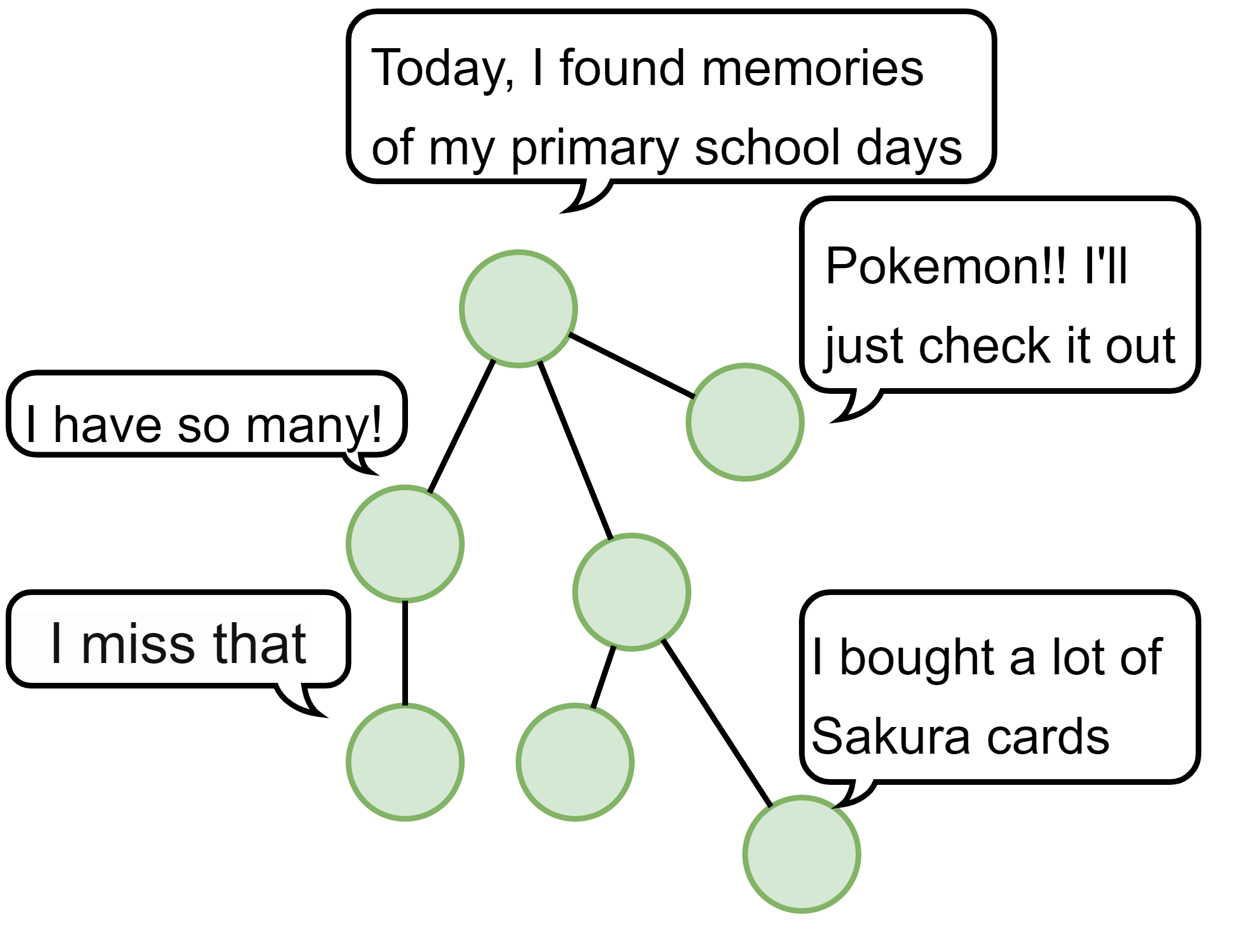}}
  \caption{Illustration examples of propagation trees.}
  \label{fig:pt}
\end{figure}

To debunk rumors, many efforts have been made. Among the existing studies, the propagation structure learning methods have made great progress with the least information and the best performance compared with other methods.
In these methods, the propagation process of a rumor is built into a tree structure, with the text features of a source post and user comments as the nodes of the tree, and the reply relationship of posts as the edges of the tree (as shown in Figure~\ref{fig:pt}). Then, the graph neural networks (GNNs) have succeeded in learning the representations of propagation trees. Typical methods all contain a carefully designed encoder or a training strategy for rumor detection with GNNs. 

Still, there are many problems in current rumor detection tasks. First, it is difficult to obtain enough labeled data to train a rumor detection model. This is mainly due to the fact that those online social media platforms supervise the release of rumors by users, and many rumors are deleted immediately after they are detected. Due to the lack of large-scale datasets, the current rumor detection methods based on rumor propagation tree structures can only be carried out on several existing common datasets, such as Weibo \cite{weibo}, Twitter15 \cite{twitter1516}, Twitter16 \cite{twitter1516}, etc. The deep learning models trained on these small size datasets usually have an overfitting problem. Second, these datasets have serious time and topic difference problems. The data in these datasets are mainly from many years ago, and the internet environment has changed a lot. Using rumor data of long time ago to detect rumors of today will lead to inconsistency between training set and testing set. What's more, newly emerged rumors are usually related to the current hot topics on the social media. It is a necessary case for making the model have semantic learning ability for recent hot topics.

Recently developed self-supervised learning (SSL) is a type of unsupervised learning method, aims to learn a robust feature representation without using explicitly provided labels as inputs, enables a model learning on unlabeled and labeled data, thereby improving the generalization of the model. 
In our scenario, although labeled rumor data is difficult to obtain, ordinary topics and their comments on a social media platform are easy to get directly, which enables us to build large-scale unlabeled ordinary topic datasets with propagation structures.
Based on the SSL techniques, we can improve the generalization performance of a model on a small-scale rumor detection dataset (labeled data) by using large-scale ordinary topics (unlabeled data). Therefore, we firstly introduce general graph SSL methods into rumor detection tasks, following a general semi-supervised graph representation learning framework without any special design for rumor detection. It should be noted that, we are not proposing some new method specific to rumor detection, but rather demonstrating that applying those general graph SSL methods can achieve comparable or even exceed the performance of those methods specifically designed for rumor detection.

We constructed two unlabeled ordinary topic datasets (named Unlabeled Weibo, UWeibo\footnote{\url{https://anonymous.4open.science/r/UWeibo-D405}} and Unlabeled Twitter, UTwitter\footnote{\url{https://anonymous.4open.science/r/UTwitter-C882}}) with massive unlabeled claims from Weibo and Twitter platform. UWeibo and UTwitter both contain approximately 200,000 claims. 
By minimizing the SSL loss of the ordinary topics in unlabeled datasets acquired in the latest time and minimizing the classification loss of the 
benchmark rumor detection datasets such as Weibo \cite{weibo}, Twitter15 \cite{twitter1516} and Twitter16 \cite{twitter1516}, we can improve the semantic learning ability and the generalization performance of a given model. 
Since benchmark datasets have been collected many years ago, there is a big time gap between them and our unlabeled datasets. We constructed a new labeled binary rumor detection dataset (named Decade Rumor Weibo, DRWeibo\footnote{\url{https://anonymous.4open.science/r/DRWeibo-16EB}}), including 6,037 claims covering a variety of topics, by integrating the rumor information of the official Weibo rumor-refuting platform\footnote{\url{https://service.account.weibo.com/?type=5}} over a decade (2012-2022).

In our experiments, we select three representative models of different types of graph SSL method 
including InfoGraph \cite{infograph}, JOAO \cite{joao} and GraphMAE \cite{graphmae}.
We adopt two commonly used training strategies, the standard pre-training and fine-tuning strategy which pre-trains on an unlabeled dataset with the SSL loss and then fine-tunes on a labeled dataset with the classification loss, the semi-supervised training strategy where SSL loss is used as the regularization term of classification loss. In addition, with the help of unlabeled data, these methods can achieve a considerable performance even with a few labeled samples. This proves that large-scale unlabeled datasets are beneficial to alleviate the overfitting and the time and topic difference problems.

In summary, the main contributions of this study are as follows.
\begin{itemize}
\item We use general graph SSL methods in rumor detection. Large-scale unlabeled dataset can be helpful for improving the models' generalization performance. 
\item We published two large-scale unlabeled topic datasets and a labeled rumor dataset, which may be helpful for rumor detection research.
\item The experimental results show that general graph SSL methods can outperform the methods specially designed for rumor detection.
\item Our few-shot experiments show that with the help of unlabeled data, the graph SSL methods can achieve good performance under few-shot scenario.
\end{itemize}

\section{RELATED WORK}

In this section, we review the related works on rumor detection and graph SSL.

\subsection{Rumor Detection}

Early studies in rumor detection used traditional classification with hand-crafted features \cite{dtc,rfc}. Deep learning's success has spawned numerous methods that significantly enhance rumor detection performance. These methods can be roughly divided into four classes, including time-series based methods \cite{yucnn,liuandwu} which model text content or user profiles as time series, propagation structure learning methods \cite{rvnn,plan,bigcn,gacl,ragcl,adgscl,pep} which consider the propagation structures of source claims and their replies, multi-source integation methods \cite{ms1,ms2} which combine multiple resources of rumors including post content, user profiles, heterogeneous relations between posts and users, multi-modal fusion methods \cite{eann,otherrumor1} which use posts' content and their related images to debunk rumors. 

Although large language models have achieved outstanding performance in many natural language processing tasks \cite{gpt3,rlhf,sspo}, propagation structure information remains crucial in rumor detection. Many state-of-the-art (SOTA) models rely on learning the representations of rumor propagation trees by GNNs. Ma et al. \cite{rvnn} constructed a bottom-up and top-down tree-structured recursive neural network to extract information from rumor propagation trees. Bian et al. \cite{bigcn} used a bidirectional GCN and the root node feature enhancement technique to classify rumors. Khoo et al. \cite{plan} constructed a Transformer that was aware of the propagation tree structure to detect rumors. Sun et al. \cite{gacl} used contrastive loss with adversarial training to learn representations that are robust to noise in rumors.

\subsection{Graph Self-Supervised Learning}

With the development of deep learning, significant progress has been made in neural message passing algorithms for supervised graph learning, achieving SOTA results for various tasks \cite{messagepass1,messagepass2,messagepass3}. However, collecting labeled data is often expensive or impossible, making graph SSL methods, which use unlabeled data to learn robust representations, increasingly popular. SSL methods include contrastive and generative approaches. Contrastive methods, initially used in image representation learning \cite{moco,simclr}, have expanded to text \cite{declutr,cert} and graph data \cite{dgi,infograph}, evolving from mutual information (MI) maximization \cite{dgi,infograph} to graph augmentation strategies \cite{mvgrl,graphcl,joao}. Adaptive augmentation \cite{gca,joao,autogcl} further refines these methods. Although less studied, generative methods like graph autoencoders \cite{gala,gate,age} offer simpler training without relying on data augmentation. GraphMAE \cite{graphmae} improves on previous autoencoders with a masked autoencoder, achieving performance comparable to SOTA contrastive methods.

\section{METHOD}

In this section, we follow a framework where any graph SSL method can be used. 

\subsection{Problem Definition}

The rumor detection task can be defined as a classification task where the goal is to predict the correct category of claims. Specifically, we denote the labeled claim set as $\mathbb{C}^{L}=\left \{c_{1},c_{2},\cdots ,c_{N_{L}}\right \}$, where $c_{i}$ represents the $i$-th claim and $N_{L}$ represents the number of labeled claims. Each labeled claim $c=(y,G)$ consists of its ground-truth label $y\in \left \{R,N\right \}$ (i.e., Rumor or Non-rumor) and its propagation graph $G=(V,E)$, where $V$ and $E$ represent the set of nodes and edges, respectively. The set of propagation structure graphs corresponding to all labeled claims is $\mathbb{G}^{L}=\left \{G_{1},G_{2},\cdots ,G_{N_{L}}\right \}$. 
The initial node feature vector matrix corresponding to $V$ is $X\in \mathbb{R}^{|V|\times d_{x}}$, and $d_{x}$ is the dimension of the initial feature vector. The initial feature vector of node can be selected widely such as Word2vec word embeddings \cite{word2vec}, tf-idf feature vectors, or feature vectors extracted from pre-trained language models \cite{bert,electra,roberta}. 
Besides the labeled dataset, we also use a large-scale unlabeled dataset. We denote the set of unlabeled claims as $\mathbb{C}^{U}=\left \{c_{N_{L}+1},c_{N_{L}+2},\cdots ,c_{N_{L}+N_{U}}\right \}$, where $N_{U}$ represents the number of unlabeled claims. There is no corresponding $y$ for unlabeled claims, but only the propagation structure $G$ for each claim. The set of propagation structure graphs corresponding to all unlabeled claims is $\mathbb{G}^{U}=\left \{G_{N_{L}+1},G_{N_{L}+2},\cdots ,G_{N_{L}+N_{U}}\right \}$. 
Our goal is to use general graph SSL methods to learn high-quality representations of graphs in $\mathbb{G}^{U}$ or even $\mathbb{G}^{L}$, simultaneously train a classifier to predict the correct categories of claims in $\mathbb{C}^{L}$.

\subsection{Framework}

\begin{figure*}[t]
  \centering
  \includegraphics[width=\textwidth]{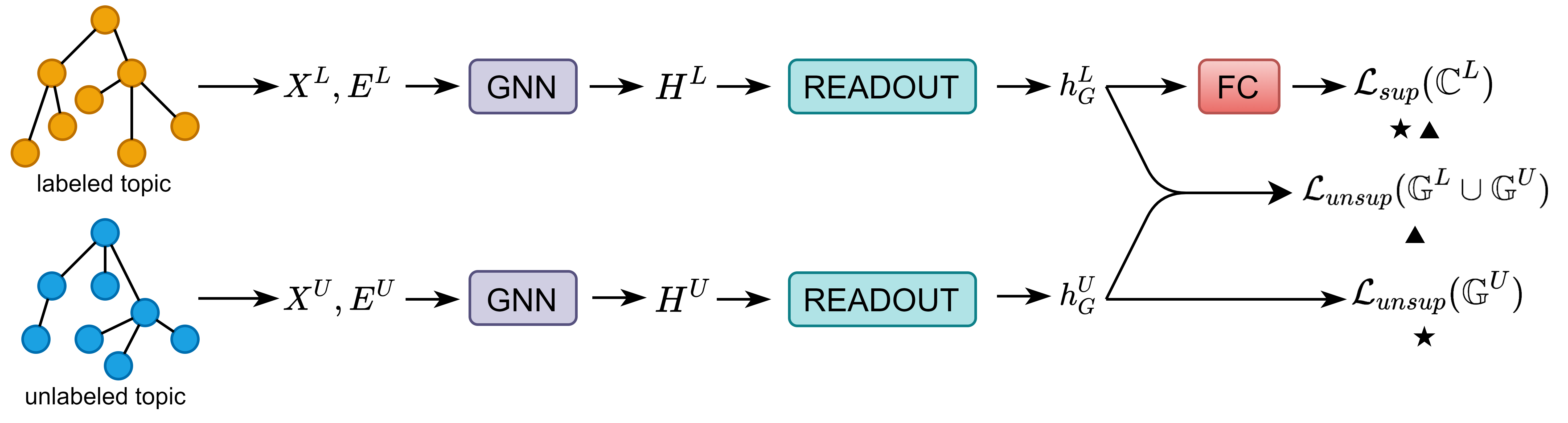}
  \caption{The general framework of graph SSL method. The $\star$ and $\blacktriangle$ represent the loss functions used by the pre-training and semi-supervised strategy.}
  \label{fig:fwk}
\end{figure*}

We view rumor detection as a general graph learning problem (see Figure~\ref{fig:fwk}). Any graph SSL method can be used. The difference of the framework from rumor detection methods based on GNNs is that we use a massive unlabeled dataset to improve the generalization ability. We conduct experiments with three representative graph SSL methods InfoGraph \cite{infograph}, JOAO \cite{joao} and GraphMAE \cite{graphmae}. These methods will calculate the unsupervised loss $\mathcal{L}_{unsup}(\mathbb{G}^{U})$ on the unlabeled graph set $\mathbb{G}^{U}$ and the unsupervised loss $\mathcal{L}_{unsup}(\mathbb{G}^{L}\cup \mathbb{G}^{U})$ on the union of labeled and unlabeled graph sets $\mathbb{G}^{L}\cup \mathbb{G}^{U}$. In addition, they need to calculate the supervised loss $\mathcal{L}_{sup}(\mathbb{C}^{L})$ on the labeled dataset $\mathbb{C}^{L}$. These losses will be applied in different combinations under the two training strategies. In order to verify the performance of general graph SSL methods for rumor detection tasks, we will not add any rumor detection specific designs to these three methods.

\subsection{InfoGraph for Rumor Detection}

InfoGraph \cite{infograph} is a graph contrastive learning method based on MI maximization, which trains an encoder of the model by maximizing the MI of the node representations and the graph representation. Supervised task and unsupervised task may prefer different information or different semantic space. Therefore, during semi-supervised learning, InfoGraph uses two encoders to alleviate the negative transfer phenomenon. In addition to minimizing the contrastive loss, it also maximizes the MI of the graph representation learned by the two encoders.

InfoGraph uses GIN \cite{gin} as an encoder $\phi$.
For a claim with propagation tree $G=(V,E)$, the node representations and graph representation are denoted as $h_{v}^{\phi }(G)$, $h_{G}^{\phi }$,
where $v\in V$. InfoGraph use Jensen-Shannon MI estimator (following the formulation of Nowozin et al. \cite{js}) to calculate the MI as below.
\begin{equation}
\begin{split}
 I(h_{v}^{\phi }(G);h_{G}^{\phi }):=\mathbb{E}_{\mathbb{P}}[-sp(-T(h_{v}^{\phi }(G),h_{G}^{\phi }))]\\-\mathbb{E}_{\mathbb{P}\times \tilde{\mathbb{P}}}[sp(T(h_{v}^{\phi }(G^{'}),h_{G}^{\phi }))],
 \end{split}
\end{equation}
where $\mathbb{P}$ is the distribution followed by $\mathbb{G}$, the set of propagation trees of a given topic dataset; $G$ is an input sample sampled from $\mathbb{P}$; $G^{'}$ is a negative sample sampled from $\tilde{\mathbb{P}}=\mathbb{P}$; $T$ is a discriminator; $sp(z)=log(1+e^{z})$ is the softplus function. In practice, we use all possible combinations of graph representations and node representations of all claim propagation graph instances within a batch to generate negative samples. The contrastive loss on $\mathbb{G}$ can be expressed as
\begin{equation}
\mathcal{L}_{unsup}(\mathbb{G})=-\frac{1}{|\mathbb{G}|}\sum _{G\in \mathbb{G}}\sum _{v\in V}I(h_{v}^{\phi }(G);h_{G}^{\phi }).
\end{equation}

When using InfoGraph for semi-supervised learning, besides $\phi$, another encoder $\varphi$ is needed. The node representations and the graph representation  of $\varphi$ are $h_{v}^{\varphi }(G)$ and $h_{G}^{\varphi }$. The MI of the graph representations by two encoders should be maximized. This is the consistency loss of two encoders on $\mathbb{G}$:
\begin{equation}
\mathcal{L}_{consistency}(\mathbb{G})=-\frac{1}{|\mathbb{G}|}\sum _{G\in \mathbb{G}}I(h_{G}^{\phi };h_{G}^{\varphi }).
\end{equation}

\textbf{Pre-Training Strategy}: InfoGraph can be pre-trained on $\mathbb{G}^{U}$ and fine-tuned on $\mathbb{G}^{L}$. In this case, one encoder is used. The loss of pre-training is
\begin{equation}
\mathcal{L}_{pretrain}=\mathcal{L}_{unsup}(\mathbb{G}^{U}).
\end{equation}
When fine-tuning, we use the pre-trained encoder to minimize classification loss:
\begin{equation}
\mathcal{L}_{fine\_tuning}=\mathcal{L}_{sup}(\mathbb{C}^{L}).
\end{equation}

\textbf{Semi-Supervised Strategy}: InfoGraph needs to use two encoders for semi-supervised learning. In this case, $\mathcal{L}_{sup}$ is the main loss, $\mathcal{L}_{unsup}$ and $\mathcal{L}_{consistency}$ are the regularization terms, then the loss of the semi-supervised strategy is
\begin{equation}
\begin{split}
\mathcal{L}_{semi}=\mathcal{L}_{sup}(\mathbb{C}^{L})+\alpha \cdot \mathcal{L}_{unsup}(\mathbb{G}^{L}\cup \mathbb{G}^{U})\\+\beta \cdot \mathcal{L}_{consistency}(\mathbb{G}^{L}\cup \mathbb{G}^{U}),
\end{split}
\end{equation}
where $\alpha$ and $\beta$ are hyperparameters. Minimizing $\mathcal{L}_{unsup}$ and $\mathcal{L}_{consistency}$ simultaneously helps the model to learn better. In practice, a batch contains both labeled and unlabeled claims, which enrich composition of negative samples and facilitate information alignment in various topics by using large-scale datasets.

\subsection{JOAO for Rumor Detection}

JOAO \cite{joao} improves the way that GraphCL \cite{graphcl} selects data augmentation. GraphCL manually selects two operations $A_{1},A_{2}$ from five different data augmentation in $\mathcal{A}=$\{NodeDrop, Subgraph, EdgePert, AttrMask, Identical\}. While JOAO binds each combination of different data augmentation with a selection probability, uses a min-max adversarial training to learn this  probability, enables to adaptively select a data augmentation stratgey to a dataset.

JOAO uses ResGCN \cite{resgcn} as an encoder $f(\cdot)$ with parameters $\theta$. Denoting the data augmentation sampling distribution by $\mathbb{P}_{(A_{1},A_{2})}$, for a set $\mathbb{G}$ of propagation trees of a dataset, its unsupervised contrastive loss is 
\begin{equation}
\begin{split}
\mathcal{L}_{unsup}(\mathbb{G})=\sum_{i=1}^{|\mathcal{A}|}\sum_{j=1}^{|\mathcal{A}|}p_{ij}\{ -\mathbb{E}_{\mathbb{P}}[sim(h_{\mathcal{A}_{i}(G)},h_{\mathcal{A}_{j}(G)})]\\+\mathbb{E}_{\mathbb{P}}[log(\mathbb{E}_{\tilde{\mathbb{P}}}exp(sim(h_{\mathcal{A}_{i}(G)},h_{\mathcal{A}_{j}(G^{'})})))] \},
\end{split}
\end{equation}
where $\mathbb{P}$ is the distribution followed by $\mathbb{G}$; $G$ is an input sample sampled from $\mathbb{P}$; $G^{'}$ is a negative sample sampled from $\tilde{\mathbb{P}}=\mathbb{P}$; $h_{\mathcal{A}_{i}(G)}$ is the graph-level representation obtained from the encoder of $\mathcal{A}_{i}(G)$ representing an augmented view of $G$; $sim(u,v)=\frac{u^{T}v}{||u||\: ||v||}$ is the cosine similarity; $p_{ij}=Prob(A_{1}=\mathcal{A}_{i},A_{2}=\mathcal{A}_{j})$.

The alternating gradient descent algorithm (AGD) \cite{agd} is used for JOAO optimization. It is divided into upper-level optimization and lower-level optimization. The upper-level optimization fixes $\mathbb{P}_{(A_{1},A_{2})}$ to minimize the contrastive loss $\mathcal{L}_{unsup}(\mathbb{G})$ to update $\theta$. The lower-level optimization fixes $\theta$ to maximize the contrastive loss $\mathcal{L}_{unsup}(\mathbb{G})$ to update $\mathbb{P}_{(A_{1},A_{2})}$, and $\mathbb{P}_{(A_{1},A_{2})}$ needs to be close to a prior distribution $\mathbb{P}_{prior}$, which is a uniform distribution making the model use multiple data augmentation combinations as much as possible. Measuring the distance between two distributions by square Euclidean distance, the distance between $\mathbb{P}_{(A_{1},A_{2})}$ and uniform distribution $\mathbb{P}_{prior}$ is $dist(\mathbb{P}_{(A_{1},A_{2})},\mathbb{P}_{prior})=\sum_{i=1}^{|\mathcal{A}|}\sum_{j=1}^{|\mathcal{A}|}(p_{ij}-\frac{1}{|\mathcal{A}|^{2}})^{2}$, and this loss can be written as
\begin{equation}
\mathcal{L}_{distribution}(\mathbb{P}_{(A_{1},A_{2})})=-\frac{1}{2}dist(\mathbb{P}_{(A_{1},A_{2})},\mathbb{P}_{prior}).
\end{equation}

\textbf{Pre-Training Strategy}: JOAO can be pre-trained on $\mathbb{G}^{U}$ and then fine-tuned on $\mathbb{G}^{L}$. In the pre-training, the upper-level and the lower-level optimization should be used alternately. The loss function of the upper-level optimization is
\begin{equation}
\mathcal{L}_{pt\_upper}=\mathcal{L}_{unsup}(\mathbb{G}^{U}).
\end{equation}
The loss function of the lower-level optimization is
\begin{equation}
\mathcal{L}_{pt\_lower}=\mathcal{L}_{unsup}(\mathbb{G}^{U})+\lambda \cdot \mathcal{L}_{distribution}(\mathbb{P}_{(A_{1},A_{2})},
\end{equation}
where $\lambda$ is an adjustable hyperparameter. When fine-tuning, it also only uses the pre-trained encoder to minimize the classification loss:
\begin{equation}
\mathcal{L}_{fine\_tuning}=\mathcal{L}_{sup}(\mathbb{C}^{L})
\end{equation}
The whole process of pre-training strategy is shown in Algorithm~\ref{alg:pre}.

\begin{algorithm}[t]
  \caption{AGD for pre-training strategy}
  \label{alg:pre}
  \begin{algorithmic}[1]
    \Require initial parameter $\theta ^{(0)}$, sampling distribution $\mathbb{P}_{(A_{1},A_{2})}^{(0)}$, pre-training step $N$, fine-tuning step $M$.
    \Ensure optimized parameter $\theta ^{(N+M)}$.
    \State $//$ Pre-training stage. 
    \For {$n=1$ to $N$}
    \State Upper-level opt: fix $\mathbb{P}_{(A_{1},A_{2})}=\mathbb{P}_{(A_{1},A_{2})}^{(n-1)}$, minimize $\mathcal{L}_{pt\_upper}$ to update $\theta ^{(n)}$.
    \State Lower-level opt: fix $\theta =\theta ^{(n)}$ and maximize $\mathcal{L}_{pt\_lower}$ to update $\mathbb{P}_{(A_{1},A_{2})}^{(n)}$.
    \EndFor
    \State $//$ Fine-tuning stage. 
    \For {$m=1$ to M}
    \State Minimize $\mathcal{L}_{fine\_tuning}$ to update $\theta ^{(N+m)}$.
    \EndFor\\
    \Return $\theta ^{(N+M)}$.
  \end{algorithmic}
\end{algorithm}

\textbf{Semi-Supervised Strategy}: JOAO needs to minimize the contrastive loss on $\mathbb{G}^{L}\cup \mathbb{G}^{U}$ when performing semi-supervised learning. The training process also follows the combination of upper-level optimization and lower-level optimization. The loss function of upper-level optimization is
\begin{equation}
\mathcal{L}_{semi\_upper}=\mathcal{L}_{sup}(\mathbb{C}^{L})+\alpha \cdot \mathcal{L}_{unsup}(\mathbb{G}^{L}\cup \mathbb{G}^{U}),
\end{equation}
where $\alpha$ is an adjustable hyperparameter. The loss of lower-level optimization is
\begin{equation}
\begin{split}
\mathcal{L}_{semi\_lower}=\mathcal{L}_{unsup}(\mathbb{G}^{L}\cup \mathbb{G}^{U})\\+\lambda \cdot \mathcal{L}_{distribution}(\mathbb{P}_{(A_{1},A_{2})}).
\end{split}
\end{equation}
The process of semi-supervised strategy is shown in Algorithm~\ref{alg:semi}.

\begin{algorithm}[t]
  \caption{AGD for semi-supervised strategy}
  \label{alg:semi}
  \begin{algorithmic}[1]
    \Require initial parameter $\theta ^{(0)}$, sampling distribution $\mathbb{P}_{(A_{1},A_{2})}^{(0)}$, optimization step $N$.
    \Ensure optimized parameter $\theta ^{(N)}$.
    \For {$n=1$ to $N$}
    \State Upper-level opt: fix $\mathbb{P}_{(A_{1},A_{2})}=\mathbb{P}_{(A_{1},A_{2})}^{(n-1)}$, minimize $\mathcal{L}_{semi\_upper}$ to update $\theta ^{(n)}$.
    \State Lower-level opt: fix $\theta =\theta ^{(n)}$, maximize $\mathcal{L}_{semi\_lower}$ to update $\mathbb{P}_{(A_{1},A_{2})}^{(n)}$.
    \EndFor\\
    \Return $\theta ^{(N)}$.
  \end{algorithmic}
\end{algorithm}

\subsection{GraphMAE for Rumor Detection}

Generative method GraphMAE \cite{graphmae} trains a masked autoencoder, whose training steps are not as complicated as a contrastive method. GraphMAE uses GCN \cite{gcn} as encoder $f_ {E}(\cdot)$ and a 1-layer GCN as decoder $f_ {D}(\cdot)$. For a propagation tree $G=(V,E)$, GraphMAE maps feature matrix $X$ to the representation matrix $H$ by $f_{E}$, then $f_{D}$ takes $H$ as input to output the reconstructed feature matrix $Z$:
\begin{equation}
H=f_{E}(X,E),Z=f_{D}(H,E).
\end{equation}

GraphMAE masks $X$ and re-masks $H$, which can effectively avoid model from learning trivial solutions. Specifically, we first sample a subset of nodes $\tilde{V}\subset V$ and mask the features of nodes in $\tilde{V}$ in $X$ with a mask token [MASK], which is a learnable vector $x_{[M]}\in \mathbb{R}^{d_{x}}$. The masked $X$ is denoted as $\tilde{X}$, and the feature vector $\tilde{x}_{v}$ of node $v\in V$ in $\tilde{X}$ is
\begin{equation}
\tilde{x}_{v}=\left\{\begin{matrix}
x_{[M]},\; \; v\in \tilde{V}\hfill \\ 
x_{v},\; \; \; \; \; \;v\notin \tilde{V}\hfill 
\end{matrix}\right..
\end{equation}

GraphMAE uses uniform distribution to randomly select nodes for masking, and the mask ratio is 50\%. In addition, the use of [MASK] will cause inconsistency between the training and inference processes. To alleviate this phenomenon, GraphMAE replaces the feature vectors of the masked nodes with the feature vectors of other nodes with a small probability (such as 15\% or less). 

Then, the representations of nodes ($\in\tilde{V}$) in $H$ are replaced by the token [DMASK], which is also a learnable vector $h_{[M]}\in \mathbb{R}^{d_{h}}$. The re-masked $H$ is denoted as $\tilde{H}$, where the feature vector $\tilde{h}_{v}$ of node $v\in V$ in $\tilde{H}$ is
\begin{equation}
\tilde{h}_{v}=\left\{\begin{matrix}
h_{[M]},\; \;v\in \tilde{V}\hfill \\ 
h_{v},\; \; \; \; \; \;v\notin \tilde{V}\hfill 
\end{matrix}\right..
\end{equation}

Taking $\tilde{H}$ as the input of $f_{D}$, the reconstructed feature matrix $Z=f_{D}(\tilde{H},E)$ is finally obtained, and the feature vector of node $v\in V$ in $Z$ is $z_{v}$. GraphMAE adopts scaled cosine error (SCE) as a loss function to avoid the sensitivity and low selectivity problems of MSE loss. GraphMAE also uses SCE as an unsupervised loss, and SCE requires the introduction of $\gamma \geq 1$ as a hyperparameter. Specifically, for a set $\mathbb{G}$ of propagation trees, the unsupervised loss is as follows.
\begin{equation}
\mathcal{L}_{unsup}(\mathbb{G})=\frac{1}{|\mathbb{G}|}\frac{1}{|\tilde{V}|}\sum _{G\in \mathbb{G}}\sum _{v\in \tilde{V}}(1-\frac{x_{v}^{T}z_{v}}{||x_{v}||\cdot ||z_{v}||})^{\gamma },\; \gamma \geq 1.
\end{equation}
This can be regarded as an adaptive weighting process for samples. The weight of each sample is adjusted with the change of reconstruction error, which is helpful for learning the semantic of a rumor with different degrees of difficulty. GraphMAE requires mask and re-mask operations only when calculating unsupervised loss, but not in calculating supervised loss and inference.

\textbf{Pre-Training Strategy}: GraphMAE is pre-trained on $\mathbb{G}^{U}$ and fine-tuned on $\mathbb{G}^{L}$. Both encoder and decoder are used in pre-training, and the loss is
\begin{equation}
\mathcal{L}_{pretrain}=\mathcal{L}_{unsup}(\mathbb{G}^{U}).
\end{equation}
Similarly, when fine-tuning, we only use the pre-trained encoder to minimize the classification loss on $\mathbb{G}^{L}$ without using decoder. The loss of fine-tuning is
\begin{equation}
\mathcal{L}_{fine\_tuning}=\mathcal{L}_{sup}(\mathbb{C}^{L}).
\end{equation}

\textbf{Semi-Supervised Strategy}: In this case, GraphMAE uses both encoder and decoder. $\mathcal{L}_{sup}$ is the main loss, with $\mathcal{L}_{unsup}$ on $\mathbb{G}^{L}$ and $\mathbb{G}^{U}$ as the regularization term:
\begin{equation}
\mathcal{L}_{semi}=\mathcal{L}_{sup}(\mathbb{C}^{L})+\alpha \cdot \mathcal{L}_{unsup}(\mathbb{G}^{L}\cup \mathbb{G}^{U}),
\end{equation}
where $\alpha$ is an adjustable hyperparameter. Minimizing $\mathcal{L}_{unsup}$ on both $\mathbb{G}^{L}$ and $\mathbb{G}^{U}$ helps the model learn high quality representations.

\section{EXPERIMENTS}

In this section, we first introduce the datasets used in the experiments and the baseline methods to which the three methods are compared. We compare the performance of InfoGraph, JOAO, GraphMAE with the baselines, and conduct ablation studies to explore the influence of unlabeled data and unsupervised loss on the models' performance. 

\subsection{Datasets}

Topics in UWeibo and UTwitter are not limited to domain, and the time period is limited to 2022 and 2023. Unlabeled data is used to enhance the learning ability of a graph model on propagation trees. Meanwhile, the varieties of topics enables a model to capture features of different topics, thus alleviating the problem of time and topic difference. In addition, we have collected the rumors from 2012 to 2022 from Weibo rumor-refuting platform and construct the new labeled dataset DRWeibo covering various topics. Table~\ref{tab:sta} shows the dataset statistics.

\begin{table*}[ht]
\caption{Statistics of the datasets.}
\centering
\begin{tabular}{ccccccc}
\Xhline{1.0pt}
\rowcolor{gray!20}
\textbf{Statistic} & \textbf{Weibo} & \textbf{DRWeibo} & \textbf{Twitter15} & \textbf{Twitter16} & \textbf{UWeibo} & \textbf{UTwitter}\\
\hline
\textbf{language} & zh & zh & en & en & zh & en \\
\textbf{labeled} & True & True & True & True & False & False \\
\textbf{\# claims} & 4664 & 6037 & 1490 & 818 & 209549 & 204922\\
\textbf{\# non-rumors} & 2351 & 3185 & 374 & 205 & - & -\\
\textbf{\# false rumors} & 2313 & 2852 & 370 & 205 & - & -\\
\textbf{\# true rumors} & - & - & 372 & 207 & - & -\\
\textbf{\# unverified rumors} & - & - & 374 & 201 & - & -\\
\textbf{\# avg posts} & 803.5 & 61.8 & 50.2 & 49.1 & 45.0 & 83.9\\
\Xhline{1.0pt}
\end{tabular}
\label{tab:sta}
\end{table*}

\subsection{Experimental Settings}

We make comparisons with the following SOTA baselines.

\textbf{PLAN} \cite{plan} uses Transformer architecture to detect rumors.

\textbf{BiGCN} \cite{bigcn} leverages two GCN encoders with different directions and root node feature enhancement strategy to classify rumor.

\textbf{UDGCN} is a variant of BiGCN, it takes only one undirected GCN encoder.

\textbf{GACL} \cite{gacl} uses contrastive learning and adversarial training to classify.

\textbf{DDGCN} \cite{ddgcn} models multiple types of information in one unified framework.

\textbf{RAGCL} \cite{ragcl} is the current SOTA method which uses contrast learning with adaptive data augmentation.

\textbf{InfoGraph, JOAO, GraphMAE} are graph SSL methods used in this work.

The baselines are re-implemented. GACL uses output of BERT \cite{bert} as initial feature vector of each post, and other models use Word2vec \cite{word2vec} word embedding trained from all posts. In experiments, Weibo and DRWeibo are used in combination with UWeibo, while Twitter15 and Twitter16 are combined with UTwitter. We evaluate accuracy (Acc.), precision (Prec.), recall (Rec.) and F1 on these datasets. The results are average results of 10 random split of the dataset. We also report the standard deviation of accuracy to reflect the stability of multiple experiment results.

\subsection{Results and Discussion}

Table~\ref{tab:weibo} and~\ref{tab:twitter} show the results on the four datasets. p and s represent pre-training and semi-supervised strategy, respectively. 
Claims in Weibo dataset contain more posts than those in other datasets, so they contain more information and it is easier for a model to learn the discriminable features of claims. Therefore, the results of all models on Weibo are significantly better than other datasets. 
Nevertheless, the improvement of the three methods on DRWeibo dataset is more obvious than other datasets, which indicates that the use of labeled and unlabeled data with a small time span is more conducive to the use of datasets with a big time gap.
In addition, the experimental results show that all the SSL methods adopted in this study outperform the models specially designed for the rumor detection, such as BiGCN and GACL. These methods are the applications of general graph SSL methods in rumor detection tasks, without any special design. Although GACL uses BERT to extract the initial feature vectors of posts, its results still fail to surpass the methods based on root node feature enhancement such as BiGCN. PLAN performs relatively poorly on the two datasets and consumes more GPU resources due to the Transformer architecture, which indicates the priority of GNN methods. 
The semi-supervised strategy is superior to the pre-training strategy most of the time, but whichever strategy is used, all models receive a benefit from the massive unlabeled datasets.

\begin{table*}[t]
\caption{Experimental results on Weibo and DRWeibo dataset.}
\centering
\begin{tabular}{cccccccccc}
 \Xhline{1.0pt}
 \rowcolor{gray!20}
 ~ & ~ & \multicolumn{4}{c}{\textbf{Weibo}} & \multicolumn{4}{c}{\textbf{DRWeibo}}\\
 \cline{3-10}
 \rowcolor{gray!20}
 \multirow{-2}{*}{\textbf{Method}} & \multirow{-2}{*}{\textbf{Class}} & \textbf{Acc.} & \textbf{Prec.} & \textbf{Rec.} & \textbf{F1} & \textbf{Acc.} & \textbf{Prec.} & \textbf{Rec.} & \textbf{F1}\\
 \hline
 \multirow{2}{*}{PLAN} & R & \multirow{2}{*}{0.915±0.007} & 0.908 & 0.923 & 0.915 & \multirow{2}{*}{0.788±0.005} & 0.786 & 0.760 & 0.771 \\
 & N & ~ & 0.923 & 0.907 & 0.914 & ~ & 0.793 & 0.813 & 0.802 \\
 \hline
 \multirow{2}{*}{BiGCN} & R & \multirow{2}{*}{0.942±0.008} & 0.919 & 0.968 & 0.942 & \multirow{2}{*}{0.866±0.010} & 0.869 & 0.849 & 0.858 \\
 & N & ~ & 0.967 & 0.918 & 0.942 & ~ & 0.863 & 0.882 & 0.872 \\
 \hline
 \multirow{2}{*}{UDGCN} & R & \multirow{2}{*}{0.940±0.007} & 0.914 & 0.971 & 0.942 & \multirow{2}{*}{0.861±0.010} & 0.839 & 0.871 & 0.855 \\
 & N & ~ & 0.969 & 0.910 & 0.938 & ~ & 0.882 & 0.852 & 0.867 \\
 \hline
 \multirow{2}{*}{GACL} & R & \multirow{2}{*}{0.938±0.006} & 0.936 & 0.940 & 0.938 & \multirow{2}{*}{0.870±0.009} & 0.865 & 0.856 & 0.860 \\
 & N & ~ & 0.940 & 0.936 & 0.938 & ~ & 0.874 & 0.882 & 0.878 \\
 \hline
 \multirow{2}{*}{DDGCN} & R & \multirow{2}{*}{0.948±0.004} & 0.924 & \textbf{0.979} & 0.951 & \multirow{2}{*}{0.878±0.005} & 0.872 & 0.864 & 0.868 \\
 & N & ~ & \textbf{0.976} & 0.917 & 0.946 & ~ & 0.883 & 0.891 & 0.887 \\
 \hline
 \multirow{2}{*}{\textbf{InfoGraph(p)}} & R & \multirow{2}{*}{0.948±0.003} & 0.925 & 0.978 & 0.950 & \multirow{2}{*}{\textbf{0.895±0.011}} & 0.893 & \textbf{0.894} & \textbf{0.893} \\
 & N & ~ & \textbf{0.976} & 0.917 & 0.946 & ~ & \textbf{0.897} & 0.896 & 0.896 \\
 \hline
 \multirow{2}{*}{\textbf{InfoGraph(s)}} & R & \multirow{2}{*}{0.953±0.005} & 0.948 & 0.957 & 0.952 & \multirow{2}{*}{0.886±0.006} & 0.877 & 0.877 & 0.876 \\
 & N & ~ & 0.958 & 0.949 & 0.953 & ~ & 0.894 & 0.893 & 0.893 \\
 \hline
 \multirow{2}{*}{\textbf{JOAO(p)}} & R & \multirow{2}{*}{0.948±0.007} & 0.933 & 0.966 & 0.949 & \multirow{2}{*}{0.873±0.005} & 0.879 & 0.841 & 0.859 \\
 & N & ~ & 0.965 & 0.930 & 0.947 & ~ & 0.867 & 0.900 & 0.883 \\
 \hline
 \multirow{2}{*}{\textbf{JOAO(s)}} & R & \multirow{2}{*}{0.951±0.004} & 0.950 & 0.952 & 0.951 & \multirow{2}{*}{0.879±0.006} & 0.860 & 0.883 & 0.871 \\
 & N & ~ & 0.952 & 0.950 & 0.951 & ~ & 0.896 & 0.875 & 0.886 \\
 \hline
 \multirow{2}{*}{\textbf{GraphMAE(p)}} & R & \multirow{2}{*}{0.944±0.004} & 0.918 & 0.977 & 0.946 & \multirow{2}{*}{0.878±0.004} & 0.873 & 0.862 & 0.867 \\
 & N & ~ & 0.974 & 0.910 & 0.941 & ~ & 0.882 & 0.891 & 0.887 \\
 \hline
 \multirow{2}{*}{\textbf{GraphMAE(s)}} & R & \multirow{2}{*}{0.948±0.004} & 0.924 & \textbf{0.979} & 0.951 & \multirow{2}{*}{0.885±0.005} & 0.874 & 0.879 & 0.876 \\
 & N & ~ & \textbf{0.976} & 0.917 & 0.946 & ~ & 0.895 & 0.890 & 0.892 \\
 \hline
 \multirow{2}{*}{RAGCL(SOTA)} & R & \multirow{2}{*}{\textbf{0.959±0.004}} & \textbf{0.953} & 0.965 & \textbf{0.959} & \multirow{2}{*}{0.894\small ±0.004} & \textbf{0.894} & 0.877 & 0.885 \\
 & N & ~ & 0.966 & \textbf{0.954} & \textbf{0.960} & ~ & 0.895 & \textbf{0.909} & \textbf{0.902} \\
 \Xhline{1.0pt}
\end{tabular}
\label{tab:weibo}
\end{table*}

\begin{table*}[t]
\caption{Experimental results on Twitter15 and Twitter16 dataset.}
\centering
\begin{tabular}{ccccccccccc}
 \Xhline{1.0pt}
 \rowcolor{gray!20}
 ~ & \multicolumn{5}{c}{\textbf{Twitter15}} & \multicolumn{5}{c}{\textbf{Twitter16}}\\
 \cline{2-11}
 \rowcolor{gray!20}
 ~ & ~ & \textbf{N} & \textbf{F} & \textbf{T} & \textbf{U} & ~ & \textbf{N} & \textbf{F} & \textbf{T} & \textbf{U}\\
 \cline{3-6}
 \cline{8-11}
 \rowcolor{gray!20}
 \multirow{-3}{*}{\textbf{Method}} & \multirow{-2}{*}{\textbf{Acc.}} & \textbf{F1} & \textbf{F1} & \textbf{F1} & \textbf{F1} & \multirow{-2}{*}{\textbf{Acc.}} & \textbf{F1} & \textbf{F1} & \textbf{F1} & \textbf{F1}\\
 \hline
 PLAN & 0.819±0.004 & 0.839 & 0.854 & 0.817 & 0.759 & 0.843±0.005 & \textbf{0.855} & 0.851 & 0.858 & 0.805 \\
 BiGCN & 0.844±0.005 & 0.856 & 0.844 & 0.863 & 0.809 & 0.880±0.009 & 0.793 & 0.912 & 0.947 & 0.849 \\
 UDGCN & 0.840±0.005 & 0.848 & 0.847 & 0.864 & 0.799 & 0.875±0.009 & 0.783 & 0.902 & 0.954 & 0.839 \\
 GACL & 0.846±0.007 & 0.859 & 0.845 & 0.865 & 0.812 & 0.891±0.004 & 0.802 & 0.929 & 0.945 & 0.872 \\
 DDGCN & 0.835±0.006 & 0.840 & 0.850 & 0.856 & 0.791 & 0.893±0.004 & 0.807 & 0.931 & 0.946 & 0.871 \\
 \hdashline
\textbf{InfoGraph(p)} & 0.843±0.005 & 0.855 & 0.851 & 0.858 & 0.805 & 0.893±0.003 & 0.804 & 0.932 & 0.947 & 0.871\\
\textbf{InfoGraph(s)} & 0.851±0.005 & 0.868 & 0.846 & \textbf{0.866} & 0.821 & 0.895±0.003 & 0.807 & 0.929 & 0.946 & \textbf{0.878}\\
\textbf{JOAO(p)} & 0.837±0.002 & 0.846 & 0.842 & 0.862 & 0.797 & 0.890±0.005 & 0.802 & 0.928 & 0.942 & 0.871\\
\textbf{JOAO(s)} & 0.848±0.007 & 0.861 & 0.847 & 0.815 & \textbf{0.866} & 0.896±0.004 & 0.811 & \textbf{0.935} & 0.951 & 0.870\\
\textbf{GraphMAE(p)} & 0.836±0.006 & 0.84 & 0.848 & 0.855 & 0.797 & 0.884±0.005 & 0.794 & 0.933 & 0.929 & 0.864\\
\textbf{GraphMAE(s)} & 0.847±0.007 & 0.859 & 0.846 & 0.865 & 0.814 & 0.892±0.003 & 0.801 & 0.932 & 0.946 & 0.871\\
\hdashline
RAGCL(SOTA) & \textbf{0.859±0.005} & \textbf{0.883} & \textbf{0.859} & 0.851 & 0.837 & \textbf{0.900±0.003} & 0.831 & 0.918 & \textbf{0.958} & 0.877 \\
 \Xhline{1.0pt}
\end{tabular}
\label{tab:twitter}
\end{table*}

\subsection{Ablation Study}

We have conducted ablation studies as shown in Table~\ref{tab:ablation}. Under the semi-supervised strategy, we test the role of SSL loss and the use of unlabeled data by observing the performance under two conditions: no unlabeled data, no SSL loss and no unlabeled data. Each model only uses the corresponding GNN for supervised classification under the condition that neither unlabeled data nor SSL loss is used (the frist result of each row). The experimental results show that on this basis, SSL loss can improve the performance of GNNs, and the models' performance can be further improved if unlabeled data is used. This shows that the applications of SSL methods to rumor detection tasks can have a certain help to the rumor classification, which proves that the three methods using SSL and unlabeled data enables to improve the generalization performance of GNNs.

\begin{table}[t]
  \caption{Ablation results on Weibo and DRWeibo}
  \centering
  \resizebox{0.49\textwidth}{!}{
  \begin{tabular}{ccccc}
    \Xhline{1.0pt}
    \rowcolor{gray!20}
    \textbf{Method} & \textbf{unsup loss} & \textbf{unlabeled data} & \textbf{Weibo} & \textbf{DRWeibo} \\
    \hline
    \multirow{3}{*}{InfoGraph} &   &   & 0.933±0.005 & 0.860±0.008 \\
    ~ & \checkmark &   & 0.939±0.006 & 0.872±0.009 \\
    ~ & \checkmark & \checkmark & 0.953±0.005 & 0.886±0.006 \\
    \hline
    \multirow{3}{*}{JOAO} &   &   & 0.935±0.004 & 0.853±0.007 \\
    ~ & \checkmark &   & 0.940±0.005 & 0.865±0.007 \\
    ~ & \checkmark & \checkmark & 0.951±0.004 & 0.879±0.006 \\
    \hline
    \multirow{3}{*}{GraphMAE} &   &   & 0.931±0.005 & 0.856±0.006 \\
    ~ & \checkmark &   & 0.938±0.006 & 0.868±0.006 \\
    ~ & \checkmark & \checkmark & 0.948±0.004 & 0.885±0.005 \\
    \Xhline{1.0pt}
  \end{tabular}
  }
  \label{tab:ablation}
\end{table}

\subsection{Few-shot Rumor Detection}

\begin{figure}[!h]
  \centering
  \includegraphics[width=8cm]{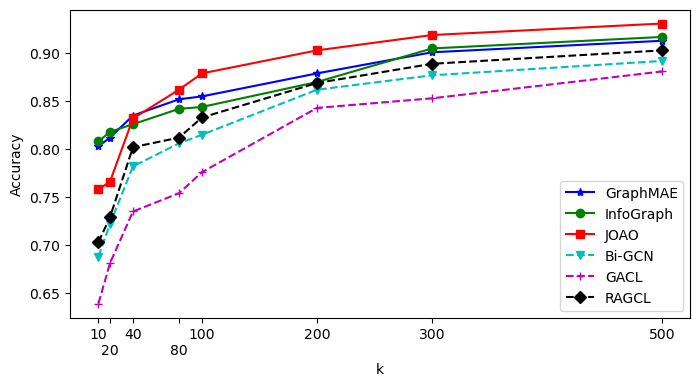}
  \caption{Results of few-shot experiments.}
  \label{fig:fs}
\end{figure}

Under the pre-training strategy, we conduct a series of few-shot experiments on Weibo dataset to explore the performance with only a small amount of labeled data (see Figure~\ref{fig:fs}). Because rumors are usually deleted after being detected, making it difficult to gain a large-scale labeled dataset, using previous rumor detection datasets to identify current time rumor topics will face the problem of time and topic difference, so the exploration of few-shot performance is very meaningful. We have conducted experiments with the number of labeled samples $k$ of 10 to 500 by pre-training on the unlabeled dataset and then fine-tuning on the labeled samples. The experimental results show that the three methods can learn better representations of claims by using unlabeled data even when there are only a few labeled samples, so as to achieve considerable performance. Even InfoGraph can achieve an accuracy of 0.808 under the condition of $k=10$.

\section{CONCLUSION}

In this study, three SSL methods are adopted for debunking rumors. Using two training strategies, we verify that applying large-scale unlabeled data of ordinary topics can improve rumor detection performance. These methods also perform well under few-shot conditions. For fulfilling our target, we have constructed two large-scale unlabeled dataset and one new rumor dataset over a decade on Weibo platform. These datasets may help to enrich the study of rumor detection field.

\section*{Acknowledgment}

This work is supported in part by the National Key R\&D Program of China (2018AAA0100302) and the National Natural Science Foundation of China (61876016).

\bibliographystyle{IEEEtran}
\bibliography{IEEEabrv}


\end{document}